\documentclass[10pt,letterpaper]{article}
\usepackage{opex3}
\usepackage{epstopdf}
\usepackage{cite}

\begin{document}

\title{H1 photonic crystal cavities for hybrid quantum information protocols}




\author{Jenna Hagemeier,$^{1,a,*}$ Cristian Bonato,$^{2,a}$ Tuan-Anh Truong,$^{3}$ Hyochul Kim,$^{1,b}$ Gareth J. Beirne,$^{2}$ Morten Bakker,$^{2}$ Martin P. van Exter,$^{2}$ Yunqiu Luo,$^{1,c}$ Pierre Petroff,$^4$ and Dirk Bouwmeester$^{1,2}$}
\address{$^1$Physics Department, University of California Santa Barbara, Santa Barbara, California 93106, USA\\
$^2$Huygens Laboratory, Leiden University, PO Box 9504, 2300 RA Leiden, The Netherlands\\
$^3$Materials Department, University of California Santa Barbara, Santa Barbara, California 93106, USA\\
$^4$ECE Department, University of California Santa Barbara, Santa Barbara, California 93106, USA\\
$^a$These authors contributed equally to this work.\\
$^b$Present address: ECE Department, IREAP, University of Maryland, College Park, Maryland 20742, USA\\
$^c$Present address: Department of Physics, Hong Kong University of Science and Technology, Clear Water Bay, Kowloon, Hong Kong, China}
\email{$^*$jenna@physics.ucsb.edu}


\begin{abstract}
Hybrid quantum information protocols are based on local qubits, such as trapped atoms, NV centers, and quantum dots, coupled to photons. The coupling is achieved through optical cavities. Here we demonstrate far-field optimized H1 photonic crystal membrane cavities combined with an additional back reflection mirror below the membrane that meet the optical requirements for implementing hybrid quantum information protocols. Using numerical optimization we find that 80$\%$ of the light can be radiated within an objective numerical aperture of 0.8, and the coupling to a single-mode fiber can be as high as 92$\%$. We experimentally prove the unique external mode matching properties by resonant reflection spectroscopy with a cavity mode visibility above 50$\%$.
\end{abstract}

\ocis{(230.5298) Photonic crystals; (270.5585) Quantum information and processing.} 

\providecommand{\noopsort}[1]{}\providecommand{\singleletter}[1]{#1}%

\section{Introduction}
Semiconductor quantum dots (QDs) coupled to optical microcavities have been extensively studied as a solid-state system for cavity quantum electrodynamics (c-QED) and as a promising platform to implement quantum information processing protocols involving single photons and single electron spins\cite{vahala2003}. Efficient single photon sources have been demonstrated\cite{orrit2005,strauf2007,forchel2010}, and the generation of polarization entangled photons has been proposed\cite{beveratos2009,hughes2009,senellart2011} and demonstrated\cite{schmidt2007,dousse2010}. Quantum information protocols employing c-QED require the implementation of a spin-photon interface, to optically read-out the spin state\cite{imamoglu2010} and to deterministically generate spin-photon entanglement\cite{rarity2008,rarity2009,bonato2010}. For these schemes and others, the cavity needs to be polarization degenerate and there needs to be good mode-matching between the input light field and the field radiated by the atom\cite{waks2006,garnier2007,vuckovic2007}. Planar photonic crystal (PhC) microcavities are widely studied because of their small mode volumes, $V \sim (\lambda/n)^3$, and high quality factors, Q $\sim10,000-50,000$\cite{noda2003}. In-plane confinement is achieved by the photonic band gap introduced by the air holes etched in a high-index membrane, while out-of-plane confinement relies on total internal reflection at the membrane-air interface. Historically, PhC cavities are not designed for good mode-matching; however, far-field optimization of PhC devices has been recently an active area of research\cite{kim2006,tran2009,tran2010,portalupi2010,pinotsi2011,haddadi2012}. In particular, the H1 defect\cite{shirane2007}, formed by removing one hole from a triangular lattice of air holes, is a promising candidate for the applications described because of its orthogonally-polarized, spectrally-degenerate dipole modes. Although fabrication imperfections lift the degeneracy of the two dipole modes, some post-fabrication techniques have been implemented to restore the degeneracy\cite{hennessy2006,fox2012}.

Here we show that H1 PhC microcavities can be optimized for high extraction efficiency while maintaining a high Q. We optimize the far-field emission patterns of our devices and suggest a compromise design in which we can achieve a collection efficiency of $80\%$ into a numerical aperture of 0.8 while maintaining a high Q of 15,000 in simulation. The maximum coupling of this cavity mode to a single mode fiber is $92\%$. We demonstrate experimentally, for the first time to our knowledge, the far-field emission pattern optimization and the improved coupling efficiency of H1 PhC devices.
This paper is organized as follows: in Sec.~\ref{sims}, we discuss our cavity design optimization and numerical simulation results; in Sec.~\ref{results}, we discuss our experimental results, including the far-field and coupling efficiency measurements; and in Sec.~\ref{conclude}, we conclude and provide an outlook of this work. 

\section{Optimized H1 cavity design}\label{sims}

\begin{figure}
\centering
\includegraphics[width=0.8\textwidth]{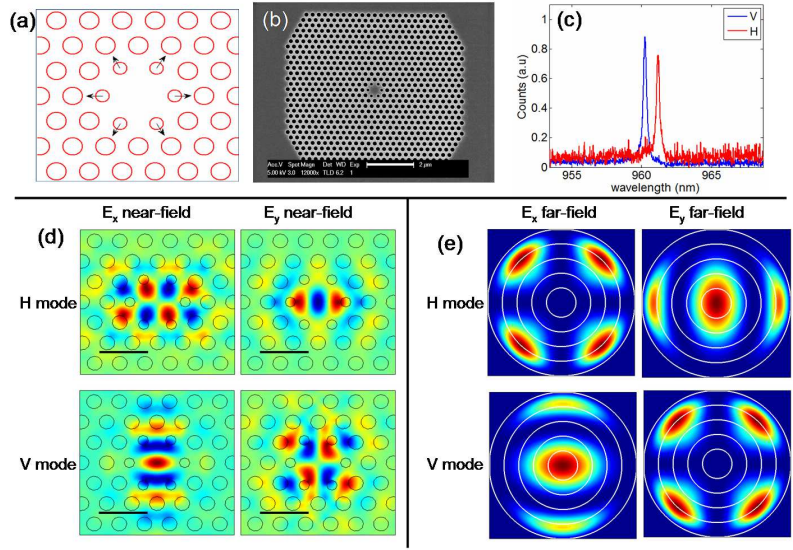}
\caption{(a) Sketch of the H1 cavity, indicating the outward shift and reduced size of the six nearest air holes. (b) SEM image of a fabricated H1 cavity PhC device; scale bar: 2 $\mu$m. (c) Characteristic photoluminescence measurement of a fabricated device, illustrating the splitting of the H and V modes due to fabrication imperfections. (d) FDTD simulated near-field amplitude components for the H dipole mode (top row) and the V dipole mode (bottom row). Plotted for each mode are Re($E_x$) and Re($E_y$). Scale bar in all plots: 250 nm. (e) FDTD simulated far-field radiation intensities of $E_x$ and $E_y$ components for the H dipole mode (top row) and the V dipole mode (bottom row). The white concentric circles correspond to NA $ = 0.2, 0.4, 0.6, 0.8,$ and $ 1.0.$}
\label{fignfff}
\end{figure}

The H1 PhC cavity design is a candidate structure for polarization degenerate light emission, and by optimizing its structural properties, we are able to dramatically improve the coupling efficiency to the cavity modes. Figure~\ref{fignfff}(a) gives a sketch of the H1 cavity, and an SEM image of a fabricated device in GaAs is shown in Fig.~\ref{fignfff}(b). The H1 cavity, being a point-like defect on a triangular lattice, has a $C_{6v}$ symmetry, i.e. the geometry remains the same rotating the structure by $\pi/3$ around the center of the cavity. In such a geometry, if a mode is an eigenstate of the wave equation, then any rotation of the mode by $\pi/3$ is also an eigenstate. If we consider a dipole-like mode, which returns to itself only under a $2\pi$ rotation, this gives six possible eigenstates, which are in general not linearly independent. It can be shown\cite{kim2003} that only two modes are linearly independent and that if $\psi_1 (r)$ is a dipole-like eigenstate, then, under the approximation of scalar fields, the orthogonal linearly-independent dipole-mode is:
\begin{equation}
\psi_2 (r) = \frac{1}{\sqrt{3}} \left[ R_{2\pi/3} \psi_1 (r) + R_{\pi/3} \psi_1 
(r) \right]
\end{equation}
where $R_x$ indicates a rotation of $x$ around the cavity center. These two degenerate orthogonally-polarized dipole-modes are perfectly suited for polarization-degenerate light emission. These considerations are valid in the ideal case of a perfect triangular lattice; in practice, however, fabrication imperfections lift the ideal structure symmetry and remove the degeneracy of the two dipole modes. This mode splitting is illustrated in the characteristic photoluminescence measurement in Fig.~\ref{fignfff}(c). These two modes, labeled H and V, each have both $x$ and $y$ E-field components, as shown in Fig.~\ref{fignfff}(d). For the H mode, $E_y$ has an antinode at the center of the cavity region, whereas the less intense $E_x$ has a node at the center, so a perfectly centered QD would primarily couple to the $E_y$ component of the H mode. In addition, the $E_x$ far-field pattern is most intense at large angles and so would be virtually undetected by a microscope objective with an NA of 0.8 or smaller, as shown in Fig.~\ref{fignfff}(e). Therefore, we can consider the H mode to be primarily $y$-polarized. For the V mode, the reverse is true, and it can therefore be considered to be primarily $x$-polarized. 

We use finite-difference-time-domain (FDTD) software\cite{fdtd} to simulate the electromagnetic field in our PhC cavities, and we perform a near- to far-field transformation\cite{vuckovic2002} using the E-field components at a plane 10 nm above the membrane surface to calculate the far-field emission profiles of the cavity modes. Simulation parameters for the near-field and far-field patterns shown in Fig.~\ref{fignfff}(d, e) are the following: lattice constant $a = 260$ nm, radius $r = 78$, and membrane thickness $d=130$ nm. In fabricated devices, the PhC membrane is separated from a Distributed-Bragg-Reflector (DBR) or the bottom substrate by an air gap. This bottom reflector, which significantly affects the far-field patterns\cite{kim2006} and improves collection efficiency of light emitted out of the bottom surface of the membrane, is modeled in our simulations as a dielectric substrate of refractive index $n=3.5$, with an air gap $L$. To improve the quality factors of the modes, as well as to improve collection efficiency through modification of the far-field profiles, optimization of the lattice structure was performed by varying the outward shift ($s$) and smaller radius ($r_{sm}$) of the six holes nearest to the cavity and by varying the air gap separation ($L$) between the membrane and the bottom reflector. While these optimization techniques have been employed by other groups to study different PhC cavity structures\cite{kim2006,beveratos2009,haddadi2012}, we combine the optimization of several parameters and demonstrate the optimization experimentally in our fabricated H1 devices. Here we will primarily discuss the role of the bottom reflector and the effect of tuning $s/a$ for the H dipole mode. A discussion of the influence of tuning $r_{sm}/r$ can be found in Appendix \ref{fdtd}. Simulation and experimental results for the orthogonally polarized V dipole mode can be found in Appendix \ref{Vmode}.

\begin{figure}
\centering
\includegraphics[width=0.55\textwidth]{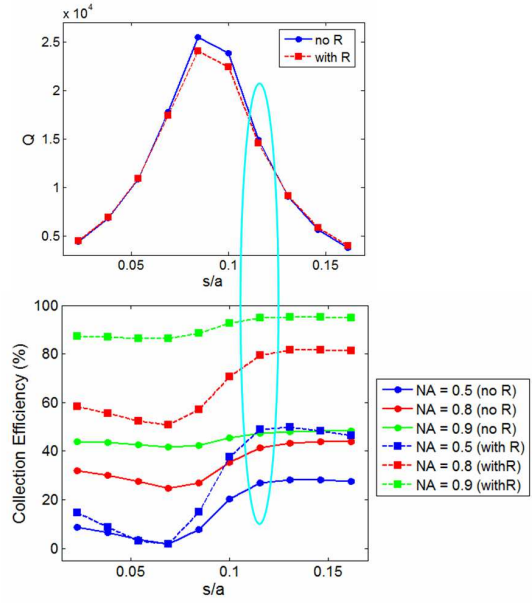}
\caption{(a) Simulated Q of the H dipole mode as a function of $s/a$. Simulations without a bottom reflector are represented by the blue circles (labeled `no R'), and simulations with a bottom reflector separated by an air gap of $L=925$ nm are represented by the red squares (labeled `with R'). (b) Simulated collection efficiency as a function of $s/a$ for NA $=$ 0.5 (in blue), 0.8 (in red), and 0.9 (in green). Simulations with (without) a bottom reflector are represented by squares (circles). For simulations without the bottom reflector, the collection efficiency maximum is 50$\%$; however, with the bottom reflector, the maximum collection efficiency is 100$\%$. The ideal compromise value, $s/a = 0.115$, is indicated by the cyan ellipse, where the Q is 15,000 and the coupling efficiency for an objective with NA of 0.8 is 80$\%$.}
\label{figQsh}
\end{figure}

From our FDTD results, we calculate coupling efficiency to a microscope objective and to a single-mode fiber. For a given far-field pattern, the collection efficiency by a microscope objective of a given NA is found by dividing the power radiated within a certain angle by the total power radiated into the half-space above the membrane. For example, a microscope objective with an NA of 0.8 will collect light emitted within an angle of $53^\circ$ from the $\Gamma$ point in k-space. For simulations without the bottom reflector, the collection efficiency maximum is 50$\%$ because only the light emitted from the top of the membrane is collected, with the other half being lost out the bottom of the membrane. In the case with the bottom reflector, however, all light can be collected above the membrane, so the maximum collection efficiency is 100$\%$. We find the single-mode fiber coupling by calculating the mode overlap of the far-field PhC mode with a Gaussian of specified mode diameter, which is the guided mode of the fiber.

The simulated quality factor and collection efficiency for cavities with different outward shift values are plotted in Fig.~\ref{figQsh}, indicating the trade-off between these figures of merit. As the collection efficiency is improved, the Q is expected to decrease as more light is being coupled to radiation modes rather than remaining confined in the dielectric slab. The collection efficiency increases for large outward shifts ($s/a$ greater than 0.1), but the Q reaches a maximum for $s/a = 0.085$. We find that $s/a=0.115$ gives a good compromise value, demonstrating an improved collection efficiency of $80\%$ for NA $= 0.8$ while still maintaining a high Q of 15,000.

\begin{figure}
\centering
\includegraphics[width=0.5\textwidth]{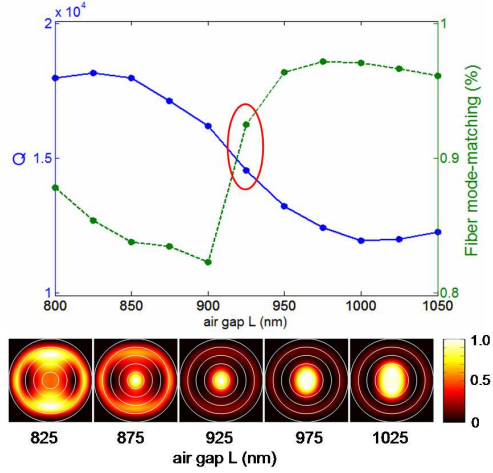}
\caption{Top: FDTD simulated Q (blue solid curve) and fiber mode-matching (green dashed curve) of the H dipole mode as a function of air gap separation $L$. The best compromise, indicated by the red ellipse, occurs for $L=925$ nm. Bottom: Simulated far-field radiation patterns for devices of different $L$ values. Each profile is normalized by its own maximum intensity, with a scale bar shown at the far right. As the air gap separation increases, the mode profiles become more Gaussian, which show better coupling to a single-mode fiber. The white concentric circles correspond to NA = 0.2, 0.4, 0.6, 0.8, and 1.0.}
\label{figair1}
\end{figure}

Figure~\ref{figQsh} compares simulations without a bottom reflector (labeled `no R') and with a bottom reflector separated by an air gap of $L=925$ nm (labeled `with R'). In general, the Q factor of the mode can be enhanced or degraded by the inclusion of a bottom reflector, which is discussed extensively in Ref.~\cite{kim2006}. The data indicate that for the specific air gap separation of 925 nm, the Q is virtually unaffected when this bottom reflector is introduced; however, the collection efficiency changes significantly. This is partially due to the fact that the collection efficiency maximum is 50$\%$ for simulations without a bottom reflector, while it can be as high as 100$\%$ when the bottom relfector is taken into account. For the optimized $s/a=0.115$ device, we performed a series of simulations to optimize the air gap separation, with results shown in Fig.~\ref{figair1}. In this case, we consider the trade-off between Q and maximum coupling to a single mode fiber, noting that the fiber mode diameter for maximum coupling may be different for each device. Far-field radiation patterns for a selection of devices are shown in the bottom panel of Fig~\ref{figair1}. We find the best compromise air gap to be $L = 925$ nm, which corresponds to $L/\lambda=0.96$. The maximum coupling between this device and a fiber mode with a beam waist of 1.5~$\mu$m is 92$\%$. The relevant parameter, $L/\lambda$ will be different for each device fabricated on a wafer; however, we chose to use $L=925$ nm for our fabricated samples.

\begin{figure}
\centering
\includegraphics[width=1.0\textwidth]{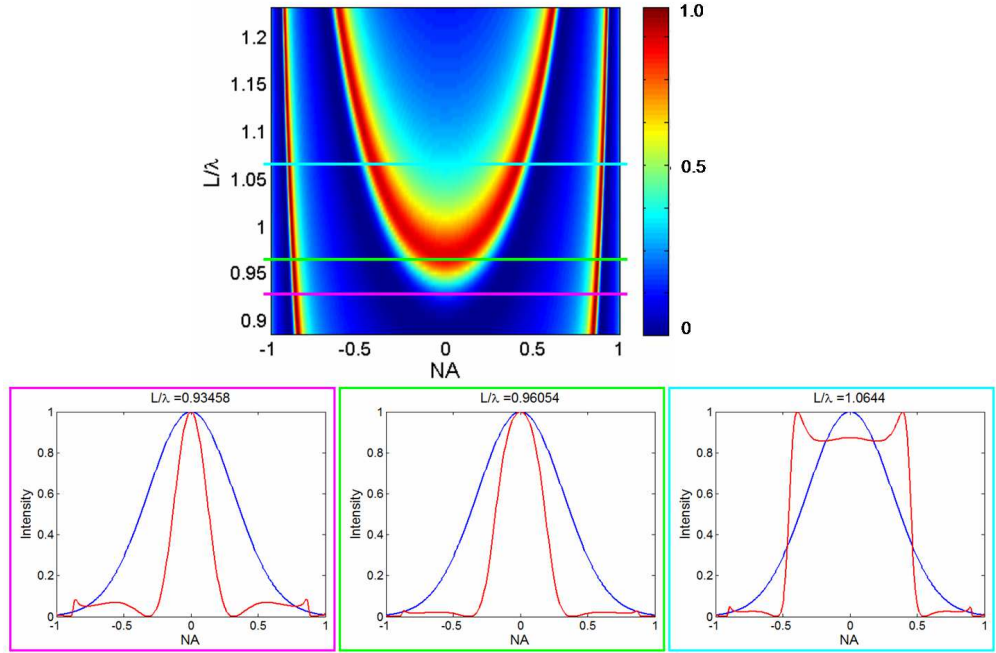}
\caption{A plot of the total radiation intensity ($W=|S+1|^2$) as a function of NA, for different values of $L/\lambda$. The calculation is done for the following paramaters: $n_{eff}=2.8$, $\lambda=963$ nm, $d=130$ nm, and $\epsilon=\pi$. In the bottom panel, line cuts are taken for $L/\lambda$ = 0.93 (left, in pink), 0.96 (middle, in green), and 1.06 (right, in cyan). Plotted are a Gaussian (blue curve) and the Gaussian convolved with the function $|S+1|^2$ (red curve). The experimental air gap of $L$ = 925 nm, for $\lambda=963$ nm, is represented by the middle green curve.}
\label{figair2}
\end{figure}

The effect of the bottom reflector on the radiation pattern can be understood by considering the interference of light that reflects multiple times between the PhC membrane and the bottom reflector. Following the treatment of Kim et al.\cite{kim2006}, let the membrane be approximated as a uniform dielectric slab with an effective refractive index, $n_{eff}$, and $L$ and $d$ are the air gap thickness and the slab thickness, respectively. Waves initially propagate upward and downward from the resonant mode of wavelength $\lambda$ in the membrane; the downward propagating wave will eventually be redirected upward and will interfere with the upward propagating wave. If we consider radiation into an angle $\theta$, the phase change through the PhC slab, $\phi$, is given by
\begin{equation}
\phi=(\frac{2\pi n_{eff}}{\lambda})d\cdot\sqrt{1-(1/n_{eff})^2 sin^2 \theta},
\end{equation}
and the round trip phase in the air gap is $2\varphi$, where
\begin{equation}
\varphi=(\frac{2\pi}{\lambda})L\cdot cos \theta.
\end{equation}
 Let $S$ be the sum of all waves detected upward which initially propagate downward. It can then be shown that\cite{kim2006}
\begin{equation}
S=\frac{t_0^2e^{i\phi}}{(1-r_0^2e^{2i\phi})(r_0+e^{-2i\varphi}e^{-i\epsilon})-r_0t_0^2e^{2i\phi}},
\end{equation}
where $\epsilon$ is the phase changes at the bottom reflector, and $r_0$ and $t_0$ are coefficients of amplitude reflection and transmission, respectively, for a single dielectric interface. An interference between $S$ and the wave initially propagating upward is characterized by $1+S$, and the resultant radiation intensity is given by
\begin{equation}
W=|1+S|^2.
\end{equation}
In Fig.~\ref{figair2}, the intensity $W$ is plotted as a function of NA and $L/\lambda$ for the following parameters: $n_{eff}=2.8$, $\lambda=963$ nm, $d=130$ nm, and $\epsilon=\pi$.

\section{Experimental results}\label{results}

\subsection{Sample fabrication}

Our samples were grown by molecular beam epitaxy (MBE) on a GaAs wafer, and included a 130 nm GaAs membrane layer with InAs QDs embedded in the middle of the membrane, a 925 nm AlGaAs sacrificial layer, and a 10 layer AlGaAs/GaAs Distributed-Bragg-Reflector (DBR) located below the sacrificial layer.  PhC H1 cavities were fabricated using the following standard recipe: e-beam lithographic patterning to define the patterns in a photoresist layer, inductively-coupled plasma dry etching to transfer the pattern into the semiconductor membrane, and hydrofluoric acid wet etching to undercut the sacrificial layer, leaving a free-standing membrane. Devices were made with systematic variations of small hole sizes and outward shifts, according to the simulated far-field optimization. Experimental Qs of our cavities were typically 3,000 to 8,000.

\subsection{Far-field imaging}

We experimentally measure the far-field profiles of our cavities by imaging the photoluminescence emission at the back focal plane of the objective onto an intensified CCD\cite{bonato2012}. We use a $2f - 2f$ imaging setup, where $f=40$ cm is the focal length of the imaging lens; our objective has an NA of 0.8. Using a CCD exposure time of 30 to 120 seconds, spectrally filtered images were collected of the far-field radiation pattern. A background image, taken with a slightly tilted interference filter, was subtracted to improve the signal-to-noise ratio.

\begin{figure}
\centering
\includegraphics[width=0.8\textwidth]{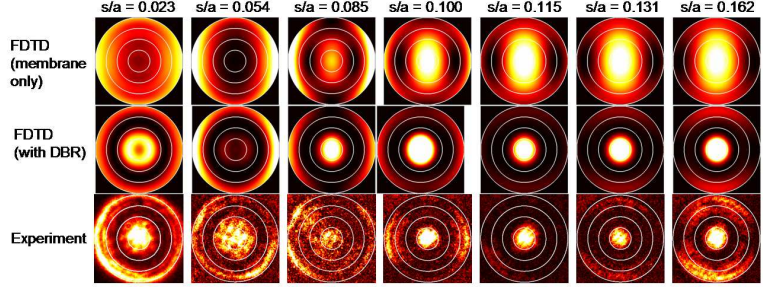}
\caption{Top row: FDTD simulated far-field profiles, considering the membrane only, for different values of $s/a$. The profiles are cut off at NA $=$ 0.8 to allow for better comparison with experimental results. Middle row: FDTD simulations, including the air gap and DBR, for the same $s/a$ values as above. For increasing shift values, the vertical beaming of the radiation clearly improves. Bottom row: Experimental far-field profiles, for one row of devices, showing a nice comparison with simulation results.}
\label{figshcomp}
\end{figure}

Figure~\ref{figshcomp} shows simulated and experimental k-space far-field profiles of the H dipole mode for different outward shift values. For small $s/a$ values, the radiation intensity is spread over a large range of angles; as $s/a$ is increased, the emitted light becomes more intense in the central region of the light cone. It is clear that the bottom reflector (middle row of Fig.~\ref{figshcomp}) modifies the far-field profiles significantly and shows much better agreement with the experimental results than the simulations of the membrane only (top row of Fig.~\ref{figshcomp}). As demonstrated in Sec.~\ref{sims}, it is sufficient to approximate the DBR below the air gap as a simple dielectric of constant refractive index for determining the ideal fabrication parameters for collection efficiency and Q factor. However, here we simulate the full 10-layer DBR structure to provide the best possible comparison with the experimental results. Small discrepancies between the experimental and simulated far-field patterns could be due to fabrication imperfections or the finite grid size of the numerical simulations. For devices with increasing outward shift values, the experimental far-field measurements show that more of the light is emitted at small angles, in good agreement with simulations.

\subsection{Resonant reflection spectroscopy measurements}

\begin{figure}
\centering
\includegraphics[width=1\textwidth]{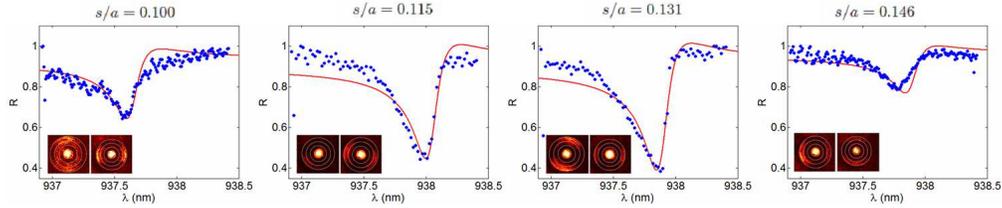}
\caption{Reflection dips for cavities with different outward shift parameters, including $s/a =$ 0.100, 0.115, 0.131, and 0.146. Data points are in blue, and the red curves are Fano lineshapes as predicted by our model, with $r_{DBR}$ = 0.95, and the mode-matching being the only free parameter. Larger dips are seen for cavities that were optimized for Gaussian-like far-field profiles. The measured far-field profiles (both H and V) for each of the cavities are shown below the reflection curves.}
\label{figfano}
\end{figure}

By modifying the far-field profiles to become Gaussian-like, the cavity modes are expected to be better mode-matched to an external input field using a microscope objective of reasonable numerical aperture. This is important in c-QED applications based on the interference between the input field and the field re-radiated by the emitter in the cavity\cite{vuckovic2007,garnier2007,rakher2009,bonato2010}. When performing measurements with a quantum dot ensemble, rather than a single quantum dot, coupled to the cavity mode, it is difficult to extract the coupling efficiency directly from the signal intensity. Instead, the improved coupling efficiency was experimentally demonstrated using resonant reflection spectroscopy. A tunable laser was scanned across the cavity resonance, and the reflected light was collected by the same microscope objective and passed through a polarizer before arriving at the detector. The polarizer was adjusted to select one of the two linearly-polarized modes.  In the case of perfect mode-matching, one would expect a dip in the reflection spectrum corresponding to the cavity resonance. In the case of quantum information schemes based on dipole-induced reflection, the bare cavity reflectivity on resonance should be ideally zero, so that the whole reflected signal can be given by the emitter coupled to the cavity. Our experimental results are shown in Fig.~\ref{figfano}. For the un-optimized cavities we could not get a detectable dip above the noise level, while for far-field optimized cavities we could get a high-contrast reflection dip, with contrast up to 60$\%$. The contrast of the dip gives a quantitative measure of the mode-matching between the cavity mode and the Gaussian laser beam input field. In this way, we demonstrate improved coupling efficiency for our far-field optimized devices. 

\begin{figure}
\centering
\includegraphics[width=0.55\textwidth]{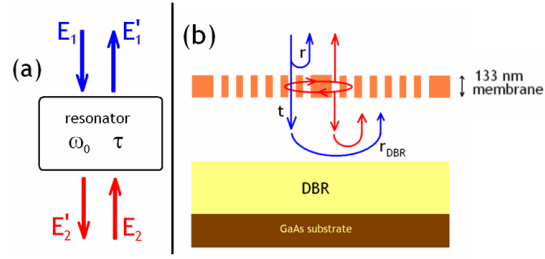}
\caption{Model for asymmetric Fano lineshapes, using a temporal coupled-mode theory. (a) Representation of the PhC cavity, with two input/output ports. (b) Schematic of the full structure, including the PhC membrane, the air gap, the DBR mirror, and the substrate.}
\label{figmodel}
\end{figure}

The reflection dips typically exhibit a Fano lineshape, which have been observed previously in PhC cavities\cite{galli2009,mccutcheon2011} and can be explained using the temporal coupled-mode theory developed by S. Fan et al.\cite{joannopoulos2003} Let us describe the photonic crystal membrane cavity as a resonator with central frequency $\omega_0$ and lifetime $\tau$, coupled to two ports, as illustrated in Fig.~\ref{figmodel}(a). In structures with mirror symmetry, the relation between the field in port-2 and port-1 is given by:
\begin{equation}\label{dip1}
\left[
\begin{array}{c}
  E'_1 \\
  E'_2
\end{array}
\right]
=
\left[
\begin{array}{cc}
  \Sigma_{11} & \Sigma_{12} \\
  \Sigma_{21} & \Sigma_{22}
\end{array}
\right]
\left[
\begin{array}{c}
  E_1 \\
  E_2
\end{array}
\right]
\end{equation}
with the following scattering matrix:
\begin{equation}\label{dip2}
\Sigma =
\left[
\begin{array}{cc}
  r & jt \\
  jt & r
\end{array}
\right]
-
\frac{(r+jt)/\tau}{j(\omega-\omega_0)+1/\tau}
\left[
\begin{array}{cc}
  1 & 1 \\
  1 & 1
\end{array}
\right]
\end{equation}
The scattering matrix is composed of two contributions. The first matrix describes the direct coupling between input and output modes, given by direct reflection or transmission by the membrane (coefficients $r$ and $t$, respectively). The second matrix, on the other hand, describes the coupling through the resonator mode. The reflectivity of the structure described by Eqs. (\ref{dip1}) and (\ref{dip2}) is:
\begin {equation}
R (\omega) = \frac{r^2 (\omega -\omega_0)^2 +(t/\tau)^2 + 2 r t (\omega-\omega_0)/\tau}{(\omega-\omega_0)^2+(1/\tau)^2}
\end{equation}
In the case of $r=0$ (or $t=0$), we just see the cavity resonance, with a Lorentzian lineshape. In all other cases, the interference between the resonance-assisted path and the path directly reflected by the membrane gives rise to a characteristic asymmetric Fano lineshape. In our case the membrane is $133$ nm thick, which is approximately half the central wavelength of the resonator. A $\lambda/2$-membrane is completely transmissive and the cavity resonance should have a Lorentzian lineshape.

However, our system is complicated by the presence of a DBR, placed at a distance $L$ from the bottom of the photonic crystal membrane (see Fig.~\ref{figmodel}(b)). The DBR couples the input and output channels for port-2:
\begin {equation}
E_2 = r_{DBR} e^{2i\omega L/c} E'_2
\end{equation}
where $r_{DBR}$ is the Fresnel coefficient of the DBR stack. Combining with $E'_2 = \Sigma_{21} E_1 + \Sigma_{22} E_2$, we get a relation between $E_2$ and $E_1$:
\begin{equation}
E_2 = \frac{r_{DBR} e^{2i\omega L/c} \Sigma_{21} (\omega)}{1-r_{DBR} e^{2i\omega L/c} \Sigma_{22} (\omega)} E_1
\end{equation}
and we can reduce the system to a one-port device with a reflectivity $|E_1'/E_1|^2$. The effective reflectivity is:
\begin{equation}
R_{eff} (\omega) = \left|
\Sigma_{11} (\omega)+
\frac{r_{DBR} e^{2i\omega L/c} \Sigma_{21} (\omega) \Sigma_{12} (\omega)}{1-r_{DBR} e^{2i\omega L/c} \Sigma_{22} (\omega)}
\right|^2
\end{equation}
The predicted reflection spectra for our experimental parameters are plotted with our data in Fig.~\ref{figfano}, with the only free parameter being the mode-matching. The predicted asymmetric reflection dips match well with our experimental observations.

\section{Conclusion}\label{conclude}

In conclusion, we have performed extensive simulations to optimize the far-field emission profiles of H1 PhC microcavities, and our experimental devices agree well with the simulation and theoretical results. We have identified PhC designs that provide excellent vertical beaming of the photoluminescence signal, as well as high contrast reflection dips up to 60$\%$. In our best compromise structure, we find that the simulated Q remains high at 15,000 and the collection efficiency is $80\%$ into a microscope objective with an NA of 0.8. The maximum mode matching of this device to a single-mode fiber is found to be 92$\%$. These results demonstrate that this platform is well-suited for realizing quantum information protocols in the solid state.

\begin{appendix}
\section{Simulation and experimental results for tuning $r_{sm}/r$}\label{fdtd}

\begin{figure}
\centering
\includegraphics[width=0.4\textwidth]{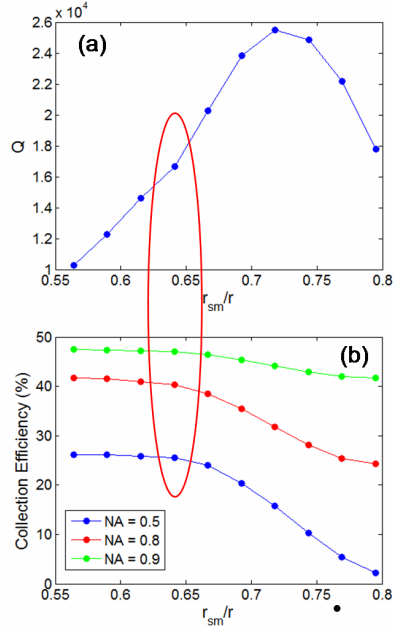}
\caption{(a) Simulated Q of the H dipole mode as a function of $r_{sm}/r$, where $r_{sm}$ represents the radius of the six nearest holes, and $r$ represents the radius of the other holes in the PhC lattice. (b) Simulated collection efficiency as a function of $r_{sm}/r$ for NA $=$ 0.5 (in blue), 0.8 (in red), and 0.9 (in green). The ideal compromise value, $r_{sm}/r=0.64$, is indicated by the red ellipse, where the Q is 17,000 and the coupling efficiency for an objective with NA of 0.8 is 40$\%$.}
\label{figQrsm}
\end{figure}

\begin{figure}
\centering
\includegraphics[width=0.7\textwidth]{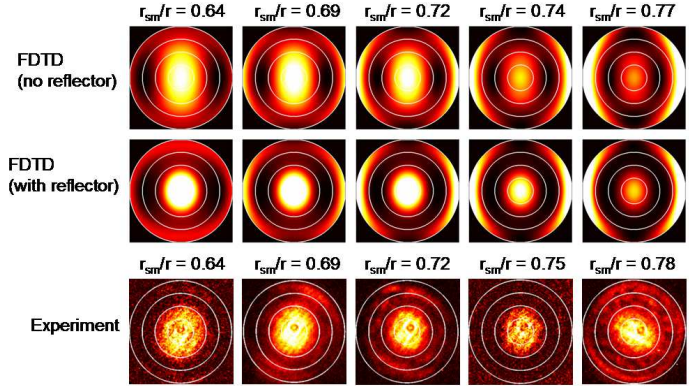}
\caption{Top row: FDTD simulated far-field profiles, considering the membrane only, for different values of r$_{sm}$/r. The profiles are cut off at NA $=$ 0.8 to allow for better comparison with experimental results. For decreasing small hole size, the radiation becomes more concentrated at small angles. Middle row: FDTD simulations, including the bottom reflector, for the same r$_{sm}$/r values as above. Bottom row: Experimental far-field profiles, for row $A$ of devices, showing a nice comparison with simulation results.} 
\label{figrsmcomp1}
\end{figure}

\begin{figure}
\centering
\includegraphics[width=0.7\textwidth]{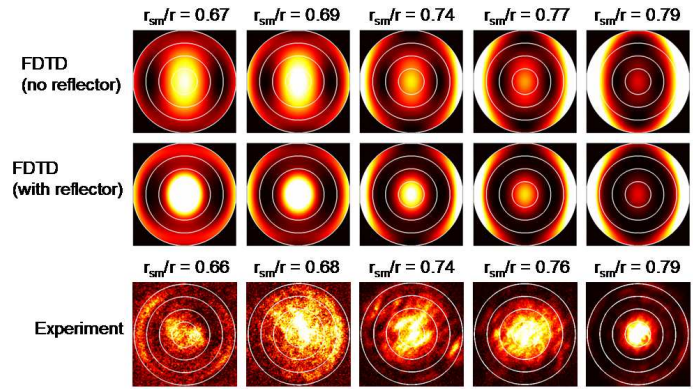}
\caption{Top row: FDTD simulated far-field profiles, considering the membrane only, for different values of r$_{sm}$/r. The profiles are cut off at NA $=$ 0.8 to allow for better comparison with experimental results. For decreasing small hole size, the radiation becomes more concentrated at small angles. Middle row: FDTD simulations, including the bottom reflector, for the same r$_{sm}$/r values as above. Bottom row: Experimental far-field profiles, for row $B$ of devices, showing a nice comparison with simulation results.} 
\label{figrsmcomp2}
\end{figure}

While the main text focused on the roles of the $s/a$ parameter and the bottom reflector in tuning the far-field profiles, we will discuss here the role of $r_{sm}/r$, which can also be varied to optimize our devices. Collection efficiency and the quality factor for simulated cavities with different $r_{sm}/r$ values are plotted in Fig.~\ref{figQrsm}, indicating the trade-off between these figures of merit. These simulation results consider only the dielectric membrane and the surrounding air, without including the bottom reflector. As such, the maximum collection efficiency is $50\%$, as only light emitted out of the top of the membrane can be collected by the objective. Collection efficiency increases for smaller $r_{sm}/r$ values, while the Q reaches a maximum for $r_{sm}/r=$ 0.72. We find that $r_{sm}/r=$ 0.64 provides a good compromise, with a collection efficiency of 40$\%$ for a microscope objective with NA $=$ 0.8 and a Q of 17,000. Simulated far-field emission patterns are shown in Fig.~\ref{figrsmcomp1} and Fig.~\ref{figrsmcomp2}. These data indicate that for small $r_{sm}/r$ values, the radiation pattern is well centered within the light cone, whereas for large $r_{sm}/r$ values, most of the light is emitted at large angles, with a less intense bright spot at the center of the light cone. The bottom reflector, approximated here as a dielectric substrate of refractive index $n=3.5$, again plays a crucial role in accurately simulating the far-field patterns. Experimentally measured far-field patterns for two rows, $A$ and $B$, of devices with different $r_{sm}/r$ values are presented in Fig.~\ref{figrsmcomp1} and Fig.~\ref{figrsmcomp2}, respectively. These data show good qualitative agreement with simulations, as the emitted light becomes more intense for small NA values as the radius of the small holes is decreased. Discrepancies between the simulated and experimental far-field patterns could be due to fabrication imperfections or the approximations made in the numerical simulations. 

\section{Optimization of the V dipole mode}\label{Vmode}
\begin{figure}
\centering
\includegraphics[width=0.7\textwidth]{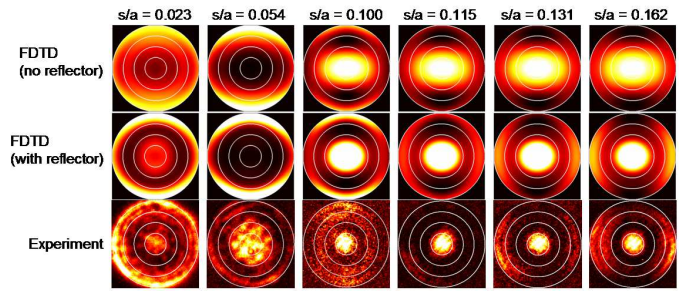}
\caption{Top row: FDTD simulated far-field profiles of the V dipole mode, considering the membrane only, for different values of $s/a$. The profiles are cut off at NA $=$ 0.8 to allow for better comparison with experimental results. For increasing shift values, the vertical beaming of the signal improves. Middle row: FDTD simulations, including the bottom reflector, for the same $s/a$ values as above. Bottom row: Experimental far-field profiles, for one row of devices, showing a nice comparison with simulation results.} 
\label{figshiftv}
\end{figure}

The FDTD simulation optimization and the experimental results for the V dipole mode are similar to those for the H dipole mode. As indicated in Fig.~\ref{figshiftv}, the far-field patterns of the V dipole mode are rotated by 90$^{\circ}$ with respect to the H dipole mode patterns. We measure experimentally the far-field profiles of the V dipole mode using the imaging procedure described in the main text. These data for devices with variation in outward shift are shown and compared with simulation in Fig.~\ref{figshiftv}. For these comparisons, the bottom reflector is approximated in the simulations as a dielectric substrate of refractive index $n=3.5$. As with the H dipole mode, the experimental far-field profiles agree well with the patterns from simulations including the bottom reflector. 

\end{appendix}

\section*{Acknowledgments}
The authors would like to thank Michiel de Dood for useful discussions. This work was supported by NSF NIRT Grant No. 0304678, Marie Curie EXT-CT-2006-042580 and FOM/NWO Grant No. 09PR2721-2. A portion of this work was carried out in the UCSB nanofabrication facility, a part of the NSF funded NNIN network.


\end{document}